\documentclass{nic-series}

\begin{document} 

\title{On the oxygen $p$ states in superconducting nickelates}

\author{Frank Lechermann}

\authortoc{F. Lechermann}

\institute{Institut f\"ur Theoretische Physik III,\\
  Ruhr-Universit\"at Bochum, D-44780 Bochum, Germany}

\maketitle

\begin{abstracts}
  While key attention in transition-metal oxides is usually devoted to the $d$ states of the
  transition-metal ion, the O$(2p)$ states usually also carry important physics. We here
  examine these $p$ states in representatives of the novel superconducting nickelates, as described
  in realistic dynamical mean-field theory. Since
  the materials are located on the boundary between Mott-Hubbard and charge-transfer systems,
  the role of oxygen is expectedly subtle. Strong reduction of doped holes on oxygen and first
  asymmetry effects are featured in infinite-layer nickelates. A pronounced nature of bridging $p_z$
  orbitals is identified in the La$_3$Ni$_2$O$_7$ system.
\end{abstracts}

\section{Introduction}
In late summer 2019, the long-sought discovery of superconductivity in nickel oxides
finally
opened up a new research field in condensed matter physics. Strontium doping of thin films of
so-called infinite-layer NdNiO$_2$ on SrTiO$_3$ substrates leads to a superconducting phase
below $T_{\rm c}\sim 15$\,K~\cite{li19}. Early follow up works revealed further similar
superconducting scenarios in Sr-doped (La,Pr)NiO$_2$, as well as in the stochiometric
multilayer compound Nd$_6$Ni$_5$O$_{10}$~\cite{pan21}. All these low-valence nickelate materials
with Ni($3d^{9-\delta}$) filling share the fact that the apical oxygens of the basic NiO$_6$
octahedra are missing.
\newline
In early spring 2024, a second class of superconducting nickelates emerged. The bilayer
La$_3$Ni$_2$O$_7$ was reported superconducting under high pressure $p>14$\,GPa with a
much higher $T_{\rm c}\sim 80$\,K~\cite{sun23}. Soon after, the trilayer compound
La$_4$Ni$_3$O$_{10}$ was proven to also show supercondcutivity in a similar pressure regime, but with an
again lower $T_{\rm c}\sim 20$\,K~\cite{zhu24}. Those nickelates with Ni($3d^{8-\delta}$) filling
have intact NiO$_6$ octahedra.
\newline
From the start, the superconducting nickelates were compared to high $T_{\rm c}$ cuprates, because
of their proximity in the periodic table and the akin building block of square-lattice
transition-metal (TM) oxide planes. However the degree of similarity in terms of physical properties
and superconducting nature is still under heavy debate. For detailed representations of the known
and discussed features from experiment and theory, we here refer to available
early review articles, e.g. Refs.~\citen{botana21,chen22,bywang24,mengwang24}.
One key issue concerns the number of relevant
Ni$(3d)$ frontier orbitals. While there is strong consensus that only the $d_{x^2-y^2}$ TM orbital is
of crucial importance in high-$T_{\rm c}$ cuprates, in general electronic properties of nickelates
the complete $e_g$ subshell $\{d_{z^2},d_{x^2-y^2}\}$ has to be taken into account. At least for the
$d^{8-\delta}$ bilayer and trilayer superconducting compounds, a pure single-orbital cuprate(-like)
physics seems very unrealistic both from theory and experiment.
\newline
Yet the present work does not focus on the detailed characteristics of the Ni$(3d)$ degrees of
freedom. Instead, it takes a deeper look on the behavior of the O$(2p)$ states in the normal state
of superconducting nickelates. While the impact of those $p$ states is more subtle, some
interesting aspects may still be revealed and learned from first-principles many body theory.
We show that the ligand-hole concept, i.e. deviation from the simplistic purely-ionic O$^{2-}$
picture, is a steady companion in these nickel oxides. It asks to be properly considered,
weighed and addressed, both in the low- and high-energy regime. Noteworthy differences in the
energetics and the occupation between $p_z$ and $p_{x,y}$ orbitals are observed. 

\section{Theoretical Background and Approach}
\label{sec_theo}
The physics of strong electron correlation in materials such as the superconducting nickelates
asks for a proper treatment of both, the realistic band theoretical aspect as provided by the
chemical bond in the solid, as well as the explicit many-body aspect of interacting electrons.
The state-of-the-art approach to realize such a faithful description is given by the hybrid
scheme of combining density functional theory (DFT) with dynamical-mean field theory (DMFT),
i.e. the so-called DFT+DMFT method (see e.g. Refs.~\citen{kotliar06} for the basics). There
usually, the transition-metal (TM) sites are the centre of attention, serving as DMFT impurities,
while the description of the ligand states remains on the DFT level (albeit surely coupled to
the strongly correlated TM sites). However in nickelates, we are most often dealing with possible
low-energy states of strongly hybridized TM$(3d)$-O$(2p)$ character. 
This is different compared to early TM oxides, e.g. titanates or vanadates, were the O$(2p)$
states only weakly hybridize into the low-energy physics and the systems reside more robustly in
the so-called Mott-Hubbard regime of correlated materials~\cite{zsa1986}.
There, the effective Hubbard $U_{\rm eff}$ is smaller than the charge-transfer energy
$\Delta=\varepsilon_d-\varepsilon_p$, with $\varepsilon_{d,p}$ describing the average onsite
energy of the TM$(3d)$ and O$(2p)$ states.
\newline
From a basic quantum-scattering perspective, TM$(3d)$ states are difficult to handle by sole band
theory. Because the $3d$ orbitals are not orthogonalized on lower lying $d$ orbitals, their electrons
approach the ion core region quite easily. This leads to the observed strong competition
between the electrons' itinerant vs. localized character in the solid state, giving rise to the
plethora of correlation effects emerging from TM$(3d)$ compounds. Notably, also O$(2p)$ is not
orthogonalized on lower lying $p$ states and therefore the electrons are also more localized than
$3p$ or even higher $p$ electrons. While due to the high electronegativity, smaller orbital
Hilbert space and strong bonding tendencies, the quantum-fluctuating character of O$(2p)$ is much
less pronounced than TM$(3d)$, correlation effects beyond DFT can still matter also from that
perspective. This may especially be true when O$(2p)$ plays a more prominent role
in the low-energy regime of late TM oxides.
Therefore on the methodological level, the differentiation in the correlation treatment of
TM$(3d)$ and O$(2p)$ states may be too severe in standard DFT+DMFT for certain classes of
materials. And understandably, issues such as the $pd$ splitting, $p$-induced renormalization
effects or intriguing ligand-mediated exchange might not be well described. There are
several ideas in going beyond this strong-differentiation treatment, for instance by combining
DMFT with the GW method~\cite{biermann03,boehnke2016,choi2016} or with screened-exchange
formalisms~\cite{vRoek14}.
Our choice builds up on introducting the self-interaction correction (SIC) scheme for oxygen
orbitals on the pseudopotential level~\cite{korner10}, to be utilized in a complete
charge-selfconsistent DFT+DMFT framework~\cite{grieger12}. This so-called
DFT+sicDMFT scheme~\cite{lechermann19} improves the $pd$-splitting description and handles
$p$-induced correlation effects (e.g. explicit oxygen-mediated band narrowing). The SIC
naturally enables correlation effects free from the need of symmetry breakings, but is
numerically much less demanding as e.g. GW+DMFT.
\newline
In the present context of superconducting nickelates, the Ni sites act as DMFT quantum impurities
and Coulomb interactions on oxygen enter by the SIC-modified pseudopotentials. The DFT part
consists of a mixed-basis pseudopotential code~\cite{elsaesser90,lechermann02,mbpp_code} and
SIC is applied to the O$(2s,2p)$ orbitals via weight factors $w_p$. While the $2s$ orbital is
fully corrected with $w_p=1.0$, the choice~\cite{korner10,lechermann19,lechermann20-1}
$w_p=0.8$ is used for $2p$ orbitals.
Continuous-time quantum Monte Carlo in hybridization-expansion scheme~\cite{werner06} as
implemented in the TRIQS code~\cite{parcollet15,seth16} solves the DMFT problem. A five-orbital
Slater-Hamiltonian, parameterized by Hubbard $U=10$\,eV and Hund exchange $J_{\rm H}=1$\,eV
\cite{lechermann20-1}, governs the correlated subspace defined by Ni$(3d)$ projected-local
orbitals~\cite{amadon08}. Crystallographic data are taken from experiment.

\section{General remarks on oxygen $p$ states in nickelates}
\label{sec_gen}
\begin{figure}[b]
\begin{center}
\includegraphics[width=\linewidth]{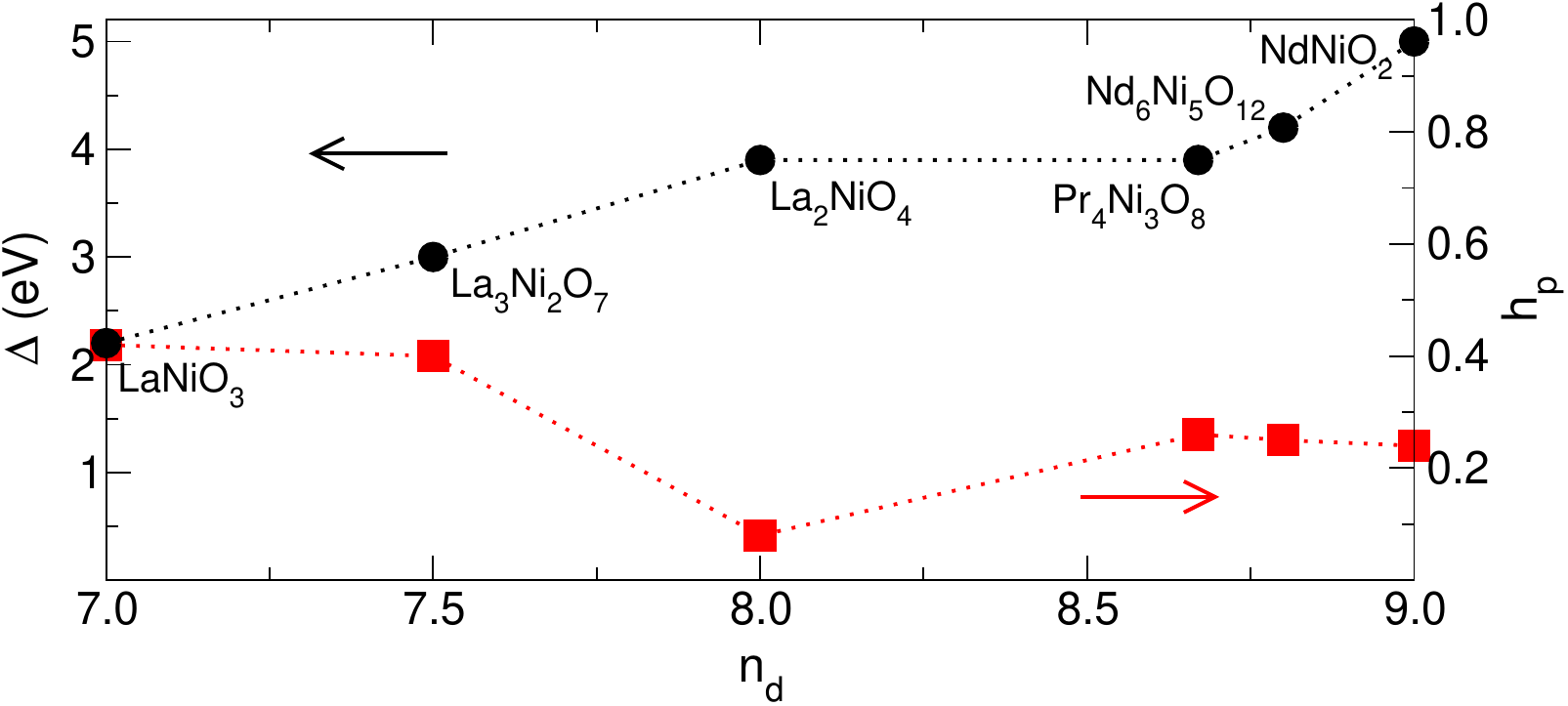}
\caption{\label{fig0}
  Charge-transfer energy $\Delta$ and O$(2p)$ hole content $h_p$ for selected
  standard ($n_d\le 8$) and reduced ($n_d>8$) Ruddlesden-Popper nickelates with
  formal Ni$(3d)$ count $n_d$. Note the plateau-like region in $\Delta$ for $8<n_d<8.67$.
}
\end{center}
\end{figure}
Nickelates are special in their correlation physics. While cuprates are usually located
in the strong charge-transfer regime, i.e., the effective Hubbard $U_{\rm eff}$ is way larger
than the charge-transfer energy $\Delta$, the nickelates are closer to a competing regime
of Mott-Hubbard versus charge-transfer dominance. Note that the calculational $U>U_{\rm eff}$
in DFT+sicDMFT, since further screening processes occur in the charge-selfconsistent many-body
approach.
The leveling of $U_{\rm eff}$ and $\Delta$ leads to a subtle role of O$(2p)$ regarding the
interplay of its low- and high-energy electronic character. In a Mott-Hubbard system, the
oxygen $p$ states reside in their localized high-energy being, whereas their itinerant
nature is most vital in a charge-transfer system. The subtlety is also reflected in the
ligand-hole physics of in fact many nickelates. Figure~\ref{fig0} displays the evolution
of the charge-transfer energy $\Delta$ and the amount of holes in the oxygen $2p$ shell (per
single O ion) $h_p$ from high to low valence of Ni in given Ruddlesden-Popper(-like) nickelates.
The data for $\Delta$ is computed from DFT+sic, i.e. a conventional Kohn-Sham calculation but
using the SIC-modified oxygen pseudopotential. As the result of the fully-interacting problem,
the data for $h_p$ is obtained from DFT+sicDMFT.
\newline
From formal Ni$^{3+}$ to formal Ni$^{+}$, the value of $\Delta$ rises more or less monotonically,
as expected from the ionic physics of these elements. This means, O$(2p)$ and Ni$(3d)$ split
energetically stronger when the Ni valence becomes smaller. The filling of the O$(2p)$ shell
is set by $\Delta$, hoppings as well as by the Coulomb interactions in the system, most notably
the on-site Coulomb $U_{pp}$ and the inter-site Coulomb $U_{pd}$. In other words, the
standard identification of O$^{2-}$ in oxides is not necessarely always true. For instance,
this ionic configuration amounts to a large Coulomb penalty within O$(2p)$, which should be
properly considered within the full group of mechanisms on the given lattice that eventually
fix $h_p$. Intuitively, we expect for small $\Delta$ a larger $h_p$ since electrons can more
easily be transfered between sites, i.e. the degree of covalency is increased. Indeed, for
formal Ni$^{3+}$ the hole content on oxygen is substantial, amounting to roughly one charge unit
within the unit cell. This ligand-hole $3d^8L$ state is well-known for nickelates with nominal
higher valence than the ideal $2+$ oxidation in the solid~\cite{bis16}.
However, as shown in Fig.~\ref{fig0}
this hole content does not fall to zero for less than nominal Ni$^{2+}$. The situation is more
intriguing due to the various electronic mechanisms at play, leading to quite some finite
covalency also for the lower-valence nickelate. Again, the O$^{2-}$ fully-ionic picture is
usually fine as a first guess, but by no means always the full truth in a complex realistic
many-body compound. This limit is here only reasonably-well reached for the
Mott-charge-transfer insulator La$_2$NiO$_4$.
\newline
While the existing basic Mott vs. charge-transfer schemes provide good principle guidance,
the actual charge distribution in a concrete material is quite sophisticated, building up on
the interplay of a multitude of physical process (in a Hilbert space enclosing usually more
than only $3d$ and $2p$ states). Boiling down the full physics to $U_{\rm eff}$ and $\Delta$
only, often appears too narrow. In this regard, also a more detailed high- vs- low-energy
differentiation governing appears of vital importance.

\section{Infinite-layer NdNiO$_2$ system}
\label{sec_inf}
\begin{figure}[t]
\begin{center}
\includegraphics[width=\linewidth]{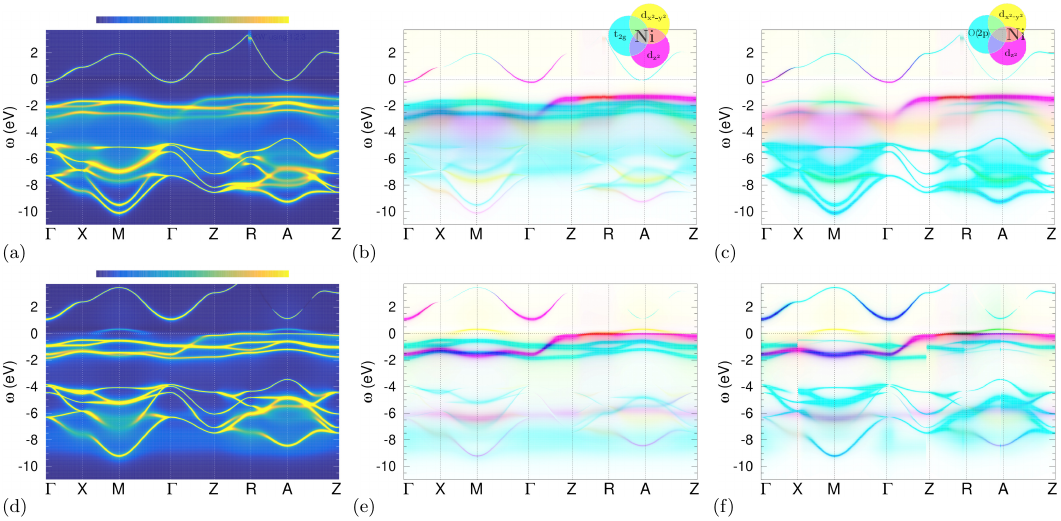}
\caption{\label{fig1}
  {\bf k}-resolved spectral function for stoichiometric NdNiO$_2$ (a-c) and with
  15\% hole doping (d-f). (a,d) full $A({\bf k},\omega)$. (b,e) Fatspec representation
  resolving Ni-$d_{x^2-y^2}$ (yellow), Ni-$d_{z^2}$ (pink) and Ni-$t_{2g}$ (cyan). (c,f) Same
  as (b,e) but Ni-$t_{2g}$ replaced by O$(2p)$.
}
\end{center}
\end{figure}

We focus on the superconducting nickelates and first analyze the O$(2p)$ states in
low-valence NdNiO$_2$ at stoichiometry and with finite hole doping. To set the stage,
Fig.~\ref{fig1} summarizes the established {\bf k}-resolved picture within
DFT+sicDMFT~\cite{lechermann20-1,lechermann21}. 
At stoichometry, Ni-$d_{x^2-y^2}$ is (nearly) Mott-insulating and finite conductivity is (mainly)
carried by weakly-filled self-doping bands (cf. Fig.~\ref{fig1}a). The latter have mixed
character of Nd$(5d)$ as well as Ni-$d_{z^2}$ (around $\Gamma$) and Ni-$d_{xz,yz}$ (around A)
(see Fig.~\ref{fig1}b,e). With hole doping, the flat-band part of Ni-$d_{z^2}$ for $k_z=1/2$ is
shifted towards the Fermi level (cf. Figs.~\ref{fig1}d), presumably playing a crucial role in
the emergence of supercoductivity. This prominent role of the Ni-$d_{z^2}$ flat-band part is
similiarly revealed in GW+DMFT~\cite{petocchi20}. Note that standard DFT+DMFT studies
without correlations on oxygen result in weaker correlations for Ni-$d_{x^2-y^2}$, enabling a more
cuprate(-like) picture for superconductivity (e.g. Refs.~\citen{kitatani20,karp20}).
\newline
Figure~\ref{fig0} reveals that even with a quite large $\Delta=5.0$\,eV, the hole content
$h_p$ amounts to 0.24. The resulting {\bf k}-resolved character of the O$(2p)$ states is depicted
in Fig.~\ref{fig1}c,f. In the stoichiometric case, O$(2p)$ dominated dispersions are easily visible
in the energy
window $\sim [-10,-5]$\,eV. With hole doping, the oxygen $p$ states mix in more strongly near
the Fermi level and into the shifted self-doping bands above. For instance around the $R$ point,
the flat band at $\varepsilon_{\rm F}$ is a strong mixture of Ni-$d_{x^2-y^2}$, Ni-$d_{z^2}$ and O$(2p)$.
Interestingly, the weak Ni-$d_{x^2-y^2}$ low-energy weight in the $k_z=0$ plane with $\Gamma$, X, M
is not sizeably hybridized with the $p$ states. This underlines the high-$\Delta$ character with
seemingly weak Zhang-Rice nature~\cite{zhangrice88} of it's low-energy physics upon
hole doping.
\begin{figure}[b]
\begin{center}
\includegraphics[width=\linewidth]{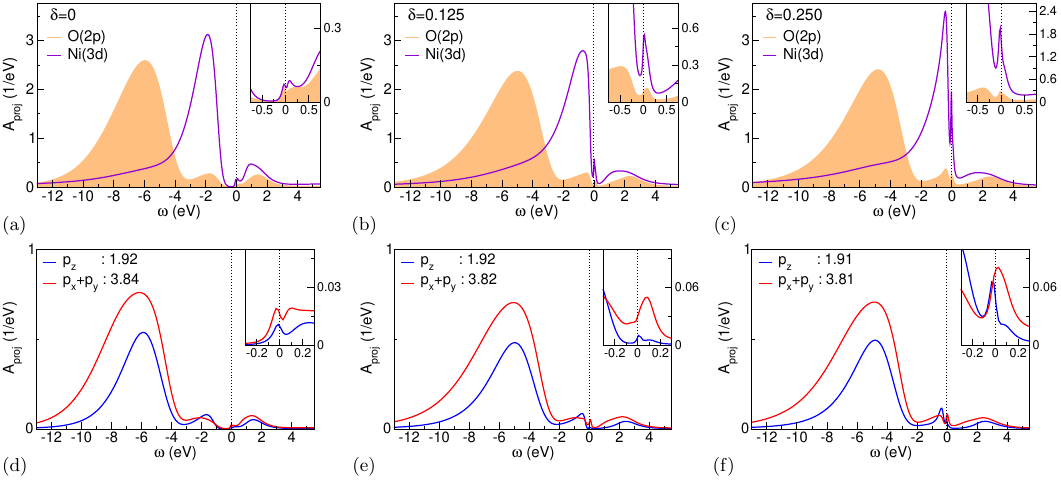}
\caption{\label{fig2}
  ${\bf k}$-integrated spectral function for stoichiometric NdNiO$_2$ (a,d), with
  12.5\% (b,e) and 25\% (c,f) hole doping. (a-c) projected O$(2p)$ and Ni$(3d)$ spectrum.
  (d-f) Projected O-$p_z$ and O-$(p_x+p_y)$ spectrum, with numbers providing the respective
  filling.
}
\end{center}
\end{figure}
\newline 
Weak Zhang-Rice character is also observable in the low-energy part of the
${\bf k}$-integrated spectra, which is displayed in a larger energy range in Figs.~\ref{fig2}(a-c).
While the small $p$ weight is comparable to the small $d$ weight at stoichiometry $\delta=0$
close to the Fermi level, the low-energy $d$ weight strongly dominates over $p$ for finite hole
doping $\delta$. Notably, as drawn from the integrated $p_z$ and $(p_x+p_y)$ spectral weight in
Figs.~\ref{fig2}(d-f), the overall oxygen hole content only increases marginally with substantial
hole doping from replacing La by Sr. In other words, most doped holes do not localize on oxygen
as in cuprates, but elsewhere, in agreement with electron energy-loss
spectroscopy~\cite{goodge21}. In the DFT+sicDMFT calculations, doped holes are mostly hosted in
the Ni-$d_{z^2}$ orbital~\cite{lechermann20-1}. Besides the total O$(2p)$ occupation, a possible
asymmetry between $p_z$ and $(p_x+p_y)$ filling may have impact~\cite{bianconi88}.
But the data shown in Figs.~\ref{fig2}(d-f) show only rather weak asymmetry for the {\sl total}
$p$ filling. Yet there is some asymmetry in the low-energy weight, especially with hole doping.
For $\delta=0.125$, the $p_z$ weight close to $\varepsilon_{\rm F}$ is quite small, while for
$\delta=0.25$ it reaches nearly the same value as the one for $(p_x+p_y)$. This may be understandable
from the special role of $p_z$, not meeting a Ni atom in nearest-neighbor distance.
For larger $\delta$, low-energy $p_z$ weight becomes more easily coherent with sizeable
Ni-$d_{z^2}$ content in the same energy range.

\section{La$_3$Ni$_2$O$_7$ system}
\label{sec_327}
\begin{figure}[t]
\begin{center}
\includegraphics[width=\linewidth]{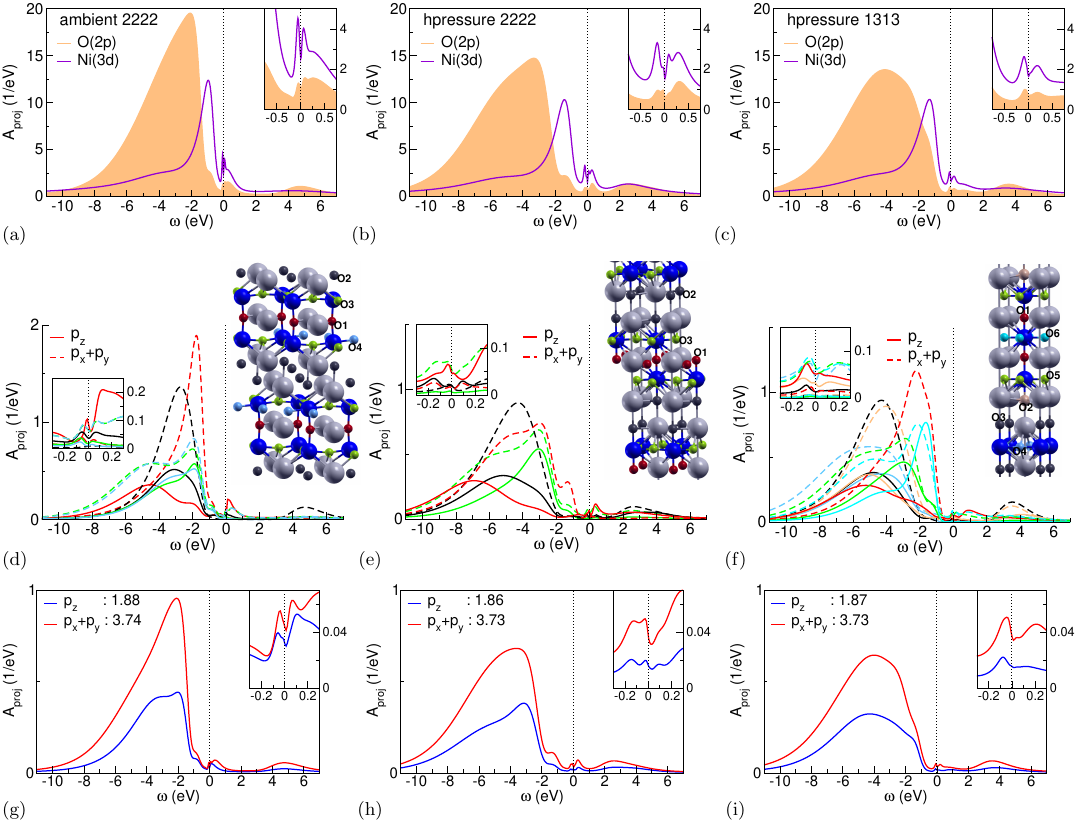}
\caption{\label{fig3}
  ${\bf k}$-integrated spectral function for La$_3$Ni$_2$O$_7$, with La: grey, Ni: darkblue.
  (a,d,g) 2222 structural motif
  at ambient pressure and (b,e,f) at high pressure, as well as (c,f,i) 1313 structural motif at
  high pressure. (a-c) Projected O$(2p)$ and Ni$(3d)$ spectrum. (d-f) Projected O-$p_z$ and
  O-$(p_x+p_y)$ spectrum for symmetry-inequivalent oxygen sites, respectively.
  (g-i) As (d-f), but summed over all oxygen sites and with numbers providing the respective
  filling. Note that possible small asymmetries between $p_{x,y}$ are averaged in the plots.
}
\end{center}
\end{figure}

The present challenges of the La$_3$Ni$_2$O$_7$ system, representing here the Ni($3d^{8-\delta}$)
superconducting-nickelate class, are not linked to missing apical oxygens or additional doping
features. Here, the sophistication lies in complexities of strucural kind and the subtle
modifications realized in the high-pressure regime. Canonically, the La$_3$Ni$_2$O$_7$ compound
is identified as the bilayer, so-called '2222', representative of the $m$-layered Ruddlesden-Popper
series La$_{m+1}$Ni$_m$O$_{3m+1}$. Recently, an alternation of mono- and trilayers, so-called
'1313', has been found as a competing structural motif~\cite{chenzhang24,puphal23,wangchenruther24}.
The distinct impact of high pressure on the electronic structure is heavily debated. At presence,
it seems that the (non-)appearance of a Ni-$d_{z^2}$-based flat band depends on pressure~\cite{mengwang24}.
\newline
Here, we again want to focus on the O$(2p)$ degrees of freedom and their main characteristics in the
correlated electronic structure. Nominally, a Ni$^{7.5}$ filling results from an ionic-limit picture
assuming O$^{2-}$ and La$^{3+}$. However, the charge-transfer energy $\Delta=3.0$ is comparatively low,
and correspondingly, the ligand-hole content $h_p=0.40$ rather high (see Fig.~\ref{fig0}). In general,
nickelates with formal oxidation state higher than Ni$^{2+}(d^8)$ form ligand holes on oxygen to keep an
effective Ni$^{2+}$ configuration. From DFT+sicDMFT, this is also true for La$_3$Ni$_2$O$_7$ with
near $d^8$ filling and one electron/hole in the $d_{z^2}$ and the $d_{x^2-y^2}$ orbital, respectively.
Recent experiments~\cite{dong24} confirm the theoretical predictions of existing ligand holes.
Figures~\ref{fig3}(a-c) shows that also the low-energy $p$-weight is somewhat enhanced compared to
the infinite-layer case, though a very strong Zhang-Rice picture as in cuprates does still not emerge.
Furthermore, applied pressure appears effective to shift both, main Ni$(3d)$ and main O$(2p)$ peak to
slightly deeper energies in the occupied spectrum.
\newline
It proves informative to plot the $p_z$- and $(p_x+p_y)$-resolved spectral weight for the different
symmetry-inequivalent oxygen sites in the given primitive cells, as done in Figs.~\ref{fig3}(d-f).
For instance, it may be observed that for all three discussed cases here, i.e. ambient-pressure 2222,
high-pressure 2222 and high-pressure 1313, the $p_{x,y}$ orbitals in the LaO flourite block separating
the NiO$_2$ multilayers are responsible for the high-energy ligand-hole peak in the unoccupied spectrum
(dashed dark curves from the dark-oxygen-ion symmetry class). One also realizes that the apical oxygens
connecting NiO$_2$ (red-colored ions), here termed bridging (BR) oxygens, show generally the
largest energy splitting between $p_z$ and $p_{x,y}$ orbitals. The most interesting behavior is
attributed to the low-energy region around the Fermi level, where the $p_z$ orbital of the BR oxygens
has apparently a prominent role. Especially for ambient-pressure 2222, the corresponding BR $p_z$ orbital
has strongest and peaked spectral weight at $\varepsilon_{\rm F}$, and furthermore a dominant low-energy
ligand-hole weight up to $\sim 0.3$\,eV above the Fermi level (see Fig.~\ref{fig3}d).
Second in low-energy weight are the
in-plane $p_{x,y}$ orbitals (green- and lightblue-colored ions) of the NiO$_2$ planes, in line with
experiment~\cite{dong24}. Hence there is a quite substantial low-energy asymmetry in favor of $p_z$ within
the available O$(2p)$ states. This finding may by linked to the spin-density-wave transition in  
ambient-pressure La$_3$Ni$_2$O$_7$, which from experiment seems majorly connected to Ni-$d_{z^2}$
involvment~\cite{xiaoyangchen2024}. The $p$-asymmetry qualitatively still holds at high pressure right
at the Fermi level, but is shifted to the higher unoccpied region above $\varepsilon_{\rm F}$
(see Figs.~\ref{fig3}e,f). For the 1313 structural motif at high pressure, the $p_z$ orbital of the
BR oxygen ion connecting inner at outer layer of the trilayer segment is more low-energy dominant
than the one connecting the trilayer with the monolayer segement. When integrating over all
symmetry-inequivalent O$(2p)$ degrees of freedom in Figs.~\ref{fig3}(g-i), the discussed asymmetries are
mostly evened out, albeit an enhanced $p_z$ low-energy weight is still easily observable for
ambient-pressure 2222. A minor slight increasement of the total hole content on O$(2p)$ with pressure
may be additionally read off from the data.

\section{Concluding Remarks}
A detailed inspection of O$(2p)$ degrees of freedom proves useful and necessary for superconducting nickelates,
because of their intriguing placement between strong Mott-Hubbard and strong charge-transfer character. Mild
orbital asymmetry and lack of significant further oxygen holes upon additional doping are revealed for the
infinite-layer systems. On the other hand, a pronounced low-energy $p$-orbital asymmetry towards the
briding $p_z$ orbital is encountered in the La$_3$Ni$_2$O$_7$ system. While some of these observations sound
rather subtle, they still may have relevant impact on the low-energy superconductivity phenomenon in these
materials. Further work in weighing the relevance of the various site- and orbital sectors in these challenging
materials is required.

\section*{Acknowledgements}
The author thanks N. Poccia, I. M. Eremin and S. B\"otzel for helpful discussions.
Computations were performed at the Ruhr-University Bochum and the JUWELS 
Cluster of the J\"ulich Supercomputing Centre (JSC) under project miqs.

\bibliography{literatur}

\end{document}